\documentclass[12pt]{article}
\usepackage{epsfig}
\usepackage{amssymb}
\DeclareFontEncoding{OT2}{}{}
\newcommand{\cyr}{\fontencoding{OT2}\fontfamily{cmr}\selectfont}
\newcommand{\singlespace}
  {\renewcommand{\baselinestretch}{1}\large\normalsize}
\newcommand{\doublespace}
  {\renewcommand{\baselinestretch}{1.6}\large\normalsize}
\renewcommand{\Im}{\mathrm{Im}\,}
\renewcommand{\Re}{\mathrm{Re}\,}
\newcommand{\MeV}{\;\mathrm{MeV}}
\newcommand{\GeV}{\;\mathrm{GeV}}
\newcommand{\fm}{\;\mathrm{fm}}
\newcommand{\abs}[1]{\vert{#1}\vert}
\newlength{\wurzelh}
\newcommand{\wurzel}[1]{%
  \settoheight{\wurzelh}{\mbox{${#1}$}}%
  \sqrt{\rule{0mm}{\wurzelh}\smash{\mbox{${#1}$}}}}
\newcommand{\D}{\Delta}
\renewcommand{\L}{\Lambda}
\renewcommand{\S}{\Sigma}
\renewcommand{\r}{\rho}
\newcommand{\w}{\omega}
\newcommand{\kv}{\vec{k}}
\newcommand{\pv}{\vec{p}\,}
\newcommand{\qv}{\vec{q}\,}
\newcommand{\rhoslash}{\r\!\!\!/}
\newcommand{\calG}{{\mathcal G}}
\newcommand{\calL}{{\mathcal L}}
\newcommand{\calPi}{\mbox{\cyr L}}
\newcommand{\rhoB}{\varrho_B}
\newcommand{\rhoN}{\varrho_N}
\newcommand{\rhoD}{\varrho_\D}
\newcommand{\rhoo}{\varrho_0}
\newcommand{\ke}{k_0+i\varepsilon}

\newcommand{\qe}{q_0+i\varepsilon}
\newcommand{\mpi}{m_\pi}
\newcommand{\mN}{m_N}
\newcommand{\mD}{m_\D}
\newcommand{\wN}{\w_N}
\newcommand{\wD}{\w_\D}
\newcommand{\GD}{\Gamma_\D}
\newcommand{\Srpp}{\S_{\r\pi\pi}}
\newcommand{\SrB}{\S_{\r B}}
\newcommand{\SrM}{\S_{\r M}}
\newcommand{\ST}{\S_T}
\newcommand{\SrppT}{\S_{\r\pi\pi\,T}}
\newcommand{\SrBT}{\S_{\r B\,T}}
\newcommand{\SrMT}{\S_{\r M\,T}}
\newcommand{\SL}{\S_L}
\newcommand{\SrppL}{\S_{\r\pi\pi\,L}}

\newcommand{\Stot}{\S_{\mathit{tot}}}
\newcommand{\Svac}{\S^{\mathit{vac}}}
\newcommand{\Smed}{\S^{\mathit{med}}}
\newcommand{\SC}{\S^C}
\newcommand{\SCmed}{\S^{C\;\mathit{med}}}
\newcommand{\pimed}{\pi_{\mathit{med}}}

\begin{document}
\begin{titlepage}
\vspace{1.0in}
\begin{flushright}
December 1999
\end{flushright}
\vspace{1.0in}
\doublespace
\begin{center}
{\large \bf Modifications of the Rho Meson from the Virtual Pion Cloud
in Hot and Dense Matter}\\
\vskip 1.0in M. Urban$^1$, M. Buballa$^1$, R. Rapp$^2$ and J. Wambach$^1$\\
{\small
{\it 1) Inst. f. Kernphysik, TU Darmstadt,
Schlo{\ss}gartenstr. 9, 64289 Darmstadt, Germany}\\
{\it 2) Department of Physics and Astronomy, State University of New York,
Stony Brook, NY 11794-3800, U.S.A.}}\\
\end{center}
\singlespace
\vspace{2cm}
\begin{abstract}
  The modification of the $\r$-meson self-energy due to the coupling to
  in-medium pions is calculated consistently at finite baryon density and
  temperature, keeping the full 3-momentum dependence in a gauge invariant
  way. As a function of nucleon density, the $\r$-meson spectral function is
  strongly enhanced in the invariant mass region $M\lesssim 650\MeV$, while
  the maximum, i.e. the pole mass, is slightly shifted upwards. As a function
  of temperature, for fixed nucleon density, the imaginary part of the
  self-energy increases further due to Bose-enhancement. At the same time the
  mass shift from the real part becomes very large. As a consequence of these
  medium effects, the dilepton rate in the low-mass region $M\lesssim 650\MeV$
  increases strongly, while the peak at $M\approx 770\MeV$ disappears.
\end{abstract}
\end{titlepage}
\setcounter{page}{2}
%
\section{Introduction}
%
\label{Sec1}
In the last few years the spectra of dilepton pairs ($e^+e^-$ or $\mu^+\mu^-$)
emerging from (ultra-) relativistic heavy-ion collisions have been measured in
various experiments. Despite of low production rates, resulting in large
statistical errors, dileptons have the distinct advantage of reaching the
detector almost undisturbed and are therefore, in principle, ideal probes for
studying the hot and dense initial phases of the fireball~\cite{Shuryak}.

The dilepton production rate in a hot and dense medium is directly related to
the electromagnetic current-current correlation function. In the hadronic
phase of the fireball and in the low invariant-mass region, $M\lesssim 1\GeV$
($M=$ invariant mass of the dilepton pair), this correlation function is
largely saturated by the $\r$-, $\w$- and $\phi$-mesons through vector-meson
dominance~\cite{Sakurai, KrollLeeZumino}. Hence dilepton spectra contain
information about the properties (mass, width) of these mesons in the
surrounding medium, most prominently the $\r$ meson due to its relatively
large dilepton decay width (see~\cite{RappWambach1} for a recent review).

\begin{figure}[b]
\begin{center}
\epsfig{file=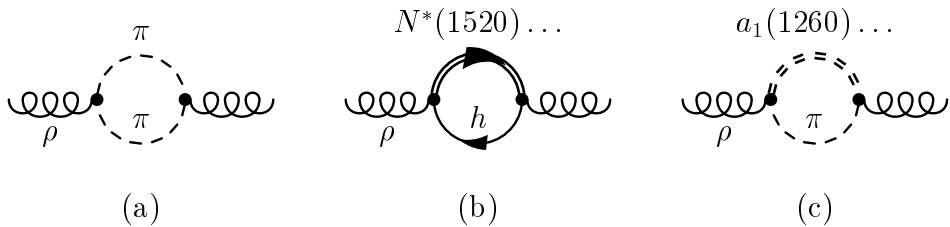}
\vspace{-.5cm}
\end{center}
\caption{\small Examples for different contributions to the $\r$-meson
  self-energy $\Stot$ in hadronic matter: (a) Decay into two pions, $\Srpp$.
  (b) Excitation of baryonic resonances via $\r N$ scattering in baryon rich
  matter, $\SrB$. (c) Excitation of mesonic resonances in $\r\pi$ scattering
  in a hot meson gas, $\SrM$.}
\label{Fig1}
\end{figure}

From a field theoretical point of view, the modifications of the $\r$-meson
properties are described by its self-energy, which contains all interactions
with the surrounding matter. In vacuum the $\r$-meson receives a width of
$\approx 150\MeV$ from the decay into two pions, corresponding to the
self-energy diagram $\Srpp$, shown in Fig.~\ref{Fig1}a.
%
%
In cold nuclear matter pions are strongly modified by Delta-hole ($\D h$)
excitations as evidenced from the wealth of pion-nucleus
data~\cite{EricsonWeise}. It is therefore suggestive to include medium
modifications by replacing the free pions in $\Srpp$, by in-medium
ones~\cite{GaleKapusta87, KorpaPratt, ChanfraySchuck, Herrmann}. Of course the
$\r$-meson can also interact resonantly with the nucleons as described by the
self-energy contributions shown in Fig.~\ref{Fig1}b and denoted by
$\SrB$~\cite{FrimanPirner, PetersPost}. At finite temperature self-energy
contributions $\SrM$, such as the one shown in Fig.~\ref{Fig1}c describing
direct interactions with thermal mesons, need to be considered in addition.
Combining these three self-energy contributions, $\Stot=\Srpp+\SrB+\SrM$, a
reasonable description of the dilepton spectra measured by the CERES
collaboration~\cite{CERES} emerges~\cite{RappChanfrayWambach, RappWambach}. In
this description the $\r$-meson acquires a substantial width and ceases to be
a well-defined quasiparticle~\cite{RappWambach1}.

In the present article we will concentrate on the contribution $\Srpp$, which
is probably the most difficult part. In the first calculations of $\Srpp$ in
matter, the flat dispersion relation of the in-medium pions produced a
divergence of $\Srpp$ near the two pion threshold~\cite{GaleKapusta87}.
However, gauge invariance was violated in these calculations. To correct this,
appropriate $\r\pi\pi$ vertex corrections had to be included, which cancel
the divergence~\cite{KorpaPratt, ChanfraySchuck, Herrmann}. Chanfray and
Schuck~\cite{ChanfraySchuck} and Herrmann et~al.~\cite{Herrmann} restricted
themselves to $\r$-mesons at rest and zero temperature. In Ref.~\cite{Urban}
we presented an extension of their models to $\r$-mesons in cold nuclear
matter with finite 3-momentum. When applied to photoabsorption cross sections
for nucleons and nuclei~\cite{RappUrban}, it was found that the pion-nucleon
and pion-Delta form factors had to be chosen rather soft, resulting in reduced
in-medium contributions from $\Srpp$.

For a consistent calculation of dilepton rates it is important to include also
temperature. This is the aim of the present article. We start with a brief
description of our model for $\Srpp$ without baryons, i.e. in vacuum or in a
hot meson gas. Section~\ref{Sec3} discusses the medium modifications of the
pion propagator in hot baryon rich matter, which will be used in
Section~\ref{Sec4} to calculate $\Srpp$ at finite temperature and baryon
density. The combined effects of temperature and baryon density on dilepton
and photon rates are studied in Section~\ref{Sec5}.
%
\section{The $\r$-meson in a hot pion gas}
\begin{figure}[b]
\begin{center}
\epsfig{file=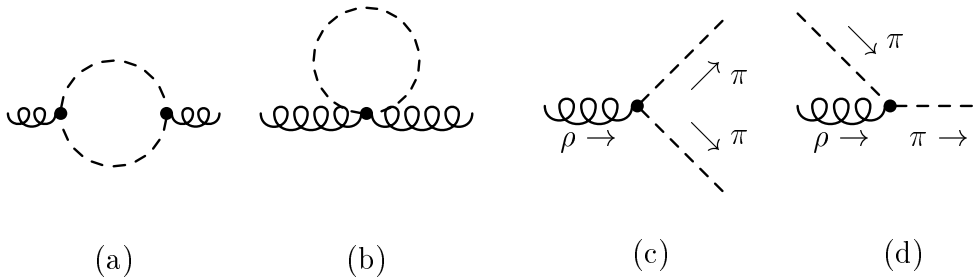}
\vspace{-.5cm}
\end{center}
\caption{\small (a) and (b) Diagrams for the $\r$-meson self-energy $\Srpp$
  at zero baryon density. (c) and (d) Processes contributing to the imaginary
  part of diagram (a) at finite temperature $T>0$: (c) $\r\rightarrow \pi+\pi$
  (decay of a $\r$-meson into two pions), (d) $\r+\pi\rightarrow \pi$
  (absorption of a virtual $\r$-meson by a thermal pion).}
\label{Fig2}
\end{figure}

%
\label{Sec2}
Before taking into account baryonic effects, we briefly discuss the $\r$-meson
self-energy at zero baryon density, $\rhoB=0$, i.e. in vacuum (temperature
$T=0$) and in a hot pion gas (temperature $T>0$). Minimal substitution leads
to the following $\pi\r$-interaction Lagrangian~\cite{Herrmann}:
\begin{equation}
\calL_{\pi\r} = \frac{1}{2} ig \r_\mu (T_3 \vec{\phi}\cdot
\partial^\mu\vec{\phi} + \partial^\mu\vec{\phi} \cdot T_3\vec{\phi})
- \frac{1}{2} g^2 \r_\mu \r^\mu T_3 \vec{\phi} \cdot T_3 \vec{\phi}\ ,
\end{equation}
where $\vec{\phi}$ denotes the isovector pion field, $\r_\mu$ the field of the
neutral $\r$-meson, $T_3$ the third component of the isospin operator and $g$
the $\r\pi\pi$ coupling constant. The $\r$-meson self-energy $\Srpp$, in the
following denoted $\S$ for simplicity, to order $g^2$ is represented by the
two diagrams shown in Figs.~\ref{Fig2}a and~\ref{Fig2}b.
%
%
Calculating these diagrams within the imaginary-time formalism described e.g.
in Ref.~\cite{FetterWalecka}, we finally obtain the following expression for
the spatial components ($i,j=1,2,3$) of the retarded $\r$-meson self-energy
tensor $\S_{\mu\nu}$ (see Section~\ref{Sec4} for details):
\begin{eqnarray}
\S_{ij}(q_0,\qv)&=&g^2 \int\frac{d^3k}{(2\pi)^3}(2k+q)_i(2k+q)_j
\Big(
  \frac{(1+2n_{\kv})(\w_{\kv+\qv}+\w_{\kv})}
    {2\w_{\kv+\qv}\w_{\kv}((\qe)^2-(\w_{\kv+\qv}+\w_{\kv})^2)}
\nonumber\\
&&\hspace{5cm}
 +\frac{n_{\kv}(\w_{\kv+\qv}-\w_{\kv})}
    {\w_{\kv+\qv}\w_{\kv}((\qe)^2-(\w_{\kv+\qv}-\w_{\kv})^2)}
\Big)\nonumber\\
&&+g^2 \delta_{ij} \int\frac{d^3k}{(2\pi)^3} \frac{1+2n_{\kv}}{\w_{\kv}}\ .
\label{Svac}\end{eqnarray}
Here we have used the short-hand notation $\w_{\kv}=\wurzel{\mpi^2+\kv^2}$
and $n_{\kv}=1/(e^{\w_{\kv}/T}-1)$.

\begin{figure}[p]
\begin{center}
\epsfig{file=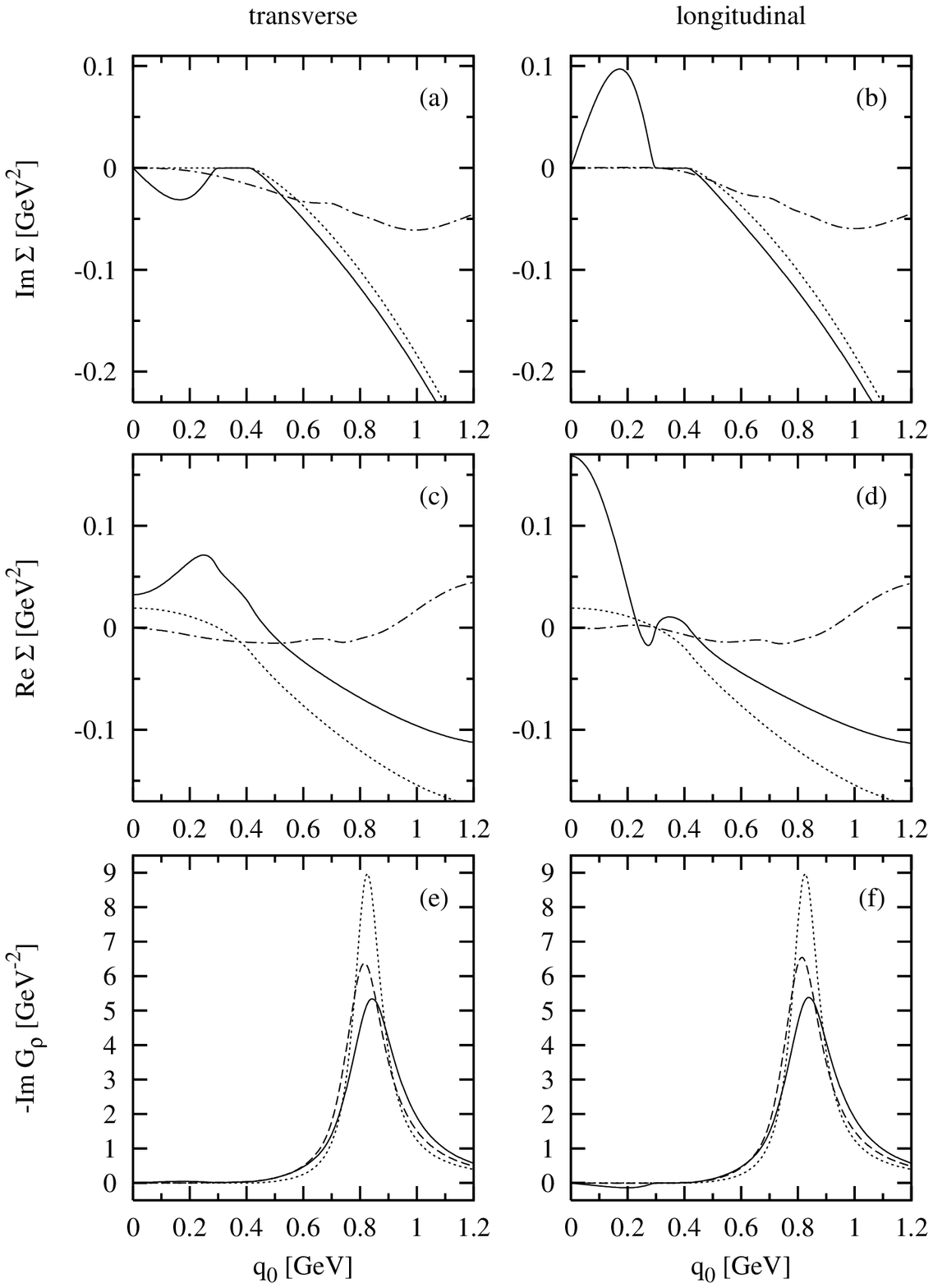}
\vspace{-.5cm}
\end{center}
\caption{\small (a) to (d) Transverse and longitudinal $\r$-meson
  self-energies for fixed 3-momentum $\abs{\qv}=300\MeV$ as functions of
  energy $q_0$, without baryons ($\rhoB=0$): $\Svac$ (dotted lines), $\Srpp$
  for $T=150\MeV$ (solid lines) and $\SrM$ for $T=150\MeV$ (dashed-dotted
  lines).  (e) and (f) Imaginary parts of the transverse and longitudinal
  $\r$-meson propagators: in vacuum (dotted lines) and for $T=150\MeV$ within
  the full model (solid lines) and neglecting temperature effects in $\Srpp$
  (dashed lines).}
\label{Fig3}
\end{figure}

The self-energy $\S$ can be separated into vacuum and medium contributions,
$\S = \Svac+\Smed$ (terms $\propto 1$ and $\propto n_{\kv}$, respectively).
In order to preserve current conservation and Lorentz invariance in the
vacuum, the divergent integrals in $\Svac$ are regularized using the
Pauli-Villars scheme. The free parameters of the model (bare $\r$ mass
$m_{\r}^{(0)} = 853\MeV$, coupling constant $g = 5.9$ and Pauli-Villars
regulator mass $\L_{\r} = 1\GeV$) are adjusted such that the electromagnetic
form factor of the pion and the $\pi\pi$-scattering phase shifts $\delta_1^1$
are reproduced satisfactorily (see Ref.~\cite{Urban} for details). Numerical
results for $\Svac$ and the corresponding $\r$-meson spectral function in
vacuum are displayed in Fig.~\ref{Fig3} (dotted lines).
%
%

Let us now discuss the self-energy at finite temperature. As there exists a
preferred frame of reference in the heat bath, $\S$ depends on $q_0$ and $\qv$
separately, and splits into 3-dimensionally transverse and longitudinal parts
$\ST$ and $\SL$~\cite{GaleKapusta91}, which are related to the spatial
components of the self-energy tensor by
\begin{equation}
\ST=\frac{1}{2} \Big(\delta^{ij}-\frac{q^i q^j}{\qv^2}\Big)\S_{ij}\ ,
\quad
\SL=\frac{q^2}{q_0^2}\frac{q^i q^j}{\qv^2}\S_{ij}\ .
\label{SijTL}
\end{equation}
(In the special case $\qv=0$ this reduces to
$\ST=\SL=\frac{1}{3}\delta^{ij}\S_{ij}$.) The angular integrations for $\ST$
and $\SL$ can be performed analytically. For time-like momenta, $q^2>0$, the
final expressions are given e.g. in Ref.~\cite{GaleKapusta91}.

Numerical results for $\SrppT$ and $\SrppL$ are shown in Figs.~\ref{Fig3}a
to~\ref{Fig3}d (solid lines). Above the two-pion threshold, $q^2>4\mpi^2$,
i.e. $q_0>410\MeV$ in Fig.~\ref{Fig3}, $\Srpp$ receives an imaginary part from
the first term in Eq.~(\ref{Svac}), which describes the decay into a
$\pi^+\pi^-$ pair (Fig.~\ref{Fig2}c). This imaginary part exists already in
vacuum, but is Bose-enhanced by a factor $1+2n_{\kv}$ at $T>0$. In the
space-like region, $q^2<0$, i.e. $q_0<300\MeV$ in Fig.~\ref{Fig3}, also the
second term in Eq.~(\ref{Svac}) generates an imaginary part in $\Srpp$,
describing the absorption of the virtual $\r$-meson by a thermal pion
(Fig.~\ref{Fig2}d) and therefore vanishing in vacuum. The main reason for the
completely different behavior of $\SrppT$ and $\SrppL$ in this region is the
factor $q^2/q_0^2$ in Eq.~(\ref{SijTL}). The real parts of $\SrppT$ and
$\SrppL$ in the time-like region are larger than the real part of $\Svac$,
thus shifting the $\r$-meson pole to higher energies. The main reasons for
this shift are the factor $1+2n_{\kv}$ in numerator of the third term in
Eq.~(\ref{Svac}) (``tadpole'' graph, Fig.~\ref{Fig2}b), and the second term,
which has a positive real part for $q_0>\qv$. In Fig.~\ref{Fig3}c, the effect
of the second term can be inferred from the steep rise of $\Re\SrppT$ in the
region where this term contributes to $\Im\SrppT$, i.e. in the interval
$0<q_0<300\MeV$.

The shift of the pole mass and the broadening of the peak due to the
Bose-enhancement of $\Im\Srpp$ can be seen more clearly in the spectral
functions.  To calculate these at finite temperature, we must consider also
the self-energy contribution $\SrM$, corresponding to diagrams like
Fig.~\ref{Fig1}c. This is not the topic of this article, so we simply take the
results from Ref.~\cite{RappGale} (dashed-dotted lines in Figs.~\ref{Fig3}a
to~\ref{Fig3}d). The spectral functions at finite temperature are now obtained
by taking the imaginary part of
\begin{equation}
G_{\r\,T,L}(q_0,\qv)=
  \frac{1}{q^2-m_{\r}^{(0)\,2}-\S_{\mathit{tot}\,T,L}}\ ,
\label{Grho}
\end{equation}
where $\Stot=\Srpp+\SrM$. These spectral functions are shown in
Figs.~\ref{Fig3}e and~\ref{Fig3}f as the solid lines. The dashed lines have
been obtained by neglecting the temperature effects in $\Srpp$, i.e. $\Stot$
in Eq.~(\ref{Grho}) has been replaced by $\Svac+\SrM$. Therefore the
difference between the dashed and the full lines is due to the temperature
effects in $\Srpp$ discussed above.
%
\section{The pion in hot nuclear matter}
%
\label{Sec3}
In cold nuclear matter the interaction of the pion with the surrounding
nucleons leads to a mixture of the pion with nucleon-hole ($Nh$) and
Delta-hole ($\D h$) excitations. In this section we will evaluate the
corresponding diagrams at finite temperature, $T$. As in our zero temperature
calculation~\cite{Urban}, we start from the interaction Lagrangians
\begin{equation}
\calL_{\pi N}=\frac{f_{\pi NN}}{\mpi}\, \bar{\psi}\gamma^5\gamma^{\mu}
  \vec{\tau}\psi\cdot \partial_{\mu}\vec{\phi}\ ,\qquad
\calL_{\pi N\D}=-\frac{f_{\pi N\D}}{\mpi}
  \bar{\psi}\vec{T}^{\dag}\psi_{\mu}\cdot\partial^{\mu}\vec{\phi}
  \; +\;\mathrm{h.c.}\ ,
\end{equation}
with $f_{\pi NN}^2/(4\pi) = 0.081$ and $f_{\pi N\D} = 2f_{\pi NN}$ (Chew-Low
value), and expand the vertices and spinors to leading order in $1/\mN$ or
$1/\mD$ to obtain standard non-relativistic Feynman rules. In the nucleon and
Delta propagators, we neglect the antiparticle contributions, but keep the
relativistic kinematics, $\w_{N,\D}(\pv) = \wurzel{m_{N,\D}^2+\pv^2}$. (We
have performed the calculations also with the non-relativistic kinematics,
$\w_{N,\D}(\pv) = m_{N,\D}+\pv^2/(2m_{N,\D})$, but found only very small
differences in the final results for the $\r$-meson self-energy.)

When calculating $Nh$ and $\D h$ excitations at finite temperature, we obtain
the following generalized Lindhard functions:
\begin{eqnarray}
\Pi_{Nh}(k_0,\kv)&=& 
  4\Big(\frac{f_{\pi NN}}{\mpi}\Big)^2\int\frac{d^3 p}{(2\pi)^3}n_N(\pv)
  \frac{2(\wN(\pv+\kv)-\wN(\pv))}{(\ke)^2-(\wN(\pv+\kv)-\wN(\pv))^2}\ ,
\\
\Pi_{\D h}(k_0,\kv)&=&
  \frac{16}{9} \Big(\frac{f_{\pi N\D}}{\mpi}\Big)^2\int\frac{d^3 p}{(2\pi)^3}
  (n_N(\pv)-n_\D(\pv+\kv)) \nonumber\\*
&&\hspace{3.5cm}\times\frac{2(\wD(\pv+\kv)-\wN(\pv))}
  {(k_0+\frac{i}{2}\GD)^2-(\wD(\pv+\kv)-\wN(\pv))^2}\ .
\label{PiDN}
\end{eqnarray}
Here we have adopted the notation
$n_{N,\D}(\pv)=1/(e^{(\w_{N,\D}(\pv)-\mu_B)/T}+1)$, and the nucleon and Delta
densities as functions of $T$ and the baryon chemical potential $\mu_B$ are
given by
\begin{equation}
\rhoN = 4 \int \frac{d^3 p}{(2\pi)^3}n_N(\pv)\ , \qquad
\rhoD = 16 \int \frac{d^3 p}{(2\pi)^3}n_\D(\pv)\ ,
\label{muB}
\end{equation}
i.e. in this context the Delta is treated as a stable particle. However, in
the $\D h$ Lindhard function $\Pi_{\D h}$ (Eq.~(\ref{PiDN})), a constant Delta
width $\GD$ is introduced, in such a way that $\Pi_{\D h}$ has the analytic
properties of a Fourier-transformed retarded function.

As described in Ref.~\cite{Urban}, we account for the effect of the repulsive
short-range $NN$ and $N\D$ interaction through phenomenological Migdal
parameters $g^\prime$~\cite{Migdal}, which is important to avoid pion
condensation. The fact that nucleon and Delta are not elementary particles is
taken into account by introducing a monopole form factor
$\Gamma_\pi(\kv)=\L^2/(\L^2+\kv^2)$ at the $\pi NN$ and $\pi N\D$ vertices.
The values of the parameters $g^\prime$ and $\L$ are constrained from a fit to
photoabsorption cross sections~\cite{Urban, RappUrban}, resulting in
$g^{\prime}_{11}=0.6$, $g^{\prime}_{12}=g^{\prime}_{22}=0.25$ and as an upper
value $\L=550\MeV$. Now the total pion self-energy reads
\begin{equation}
\Pi=\Gamma_\pi^2\,
\frac{\Pi_{Nh}+\Pi_{\D h}
  -(g^{\prime}_{11}-2g^{\prime}_{12}+g^{\prime}_{22})\Pi_{Nh}\Pi_{\D h}}
{1-g^{\prime}_{11}\Pi_{Nh}-g^{\prime}_{22}\Pi_{\D h}
  +(g^{\prime}_{11}g^{\prime}_{22}-g^{\prime 2}_{12})\Pi_{Nh}\Pi_{\D h}}
\ ,
\end{equation}
and the in-medium pion propagator is given by
\begin{equation}
G_\pi(k_0,\kv)=\frac{1}{(\ke)^2-\mpi^2-\kv^2 (1+\Pi(k_0,\kv))}\ .
\end{equation}

\begin{figure}[t]
\begin{center}
\epsfig{file=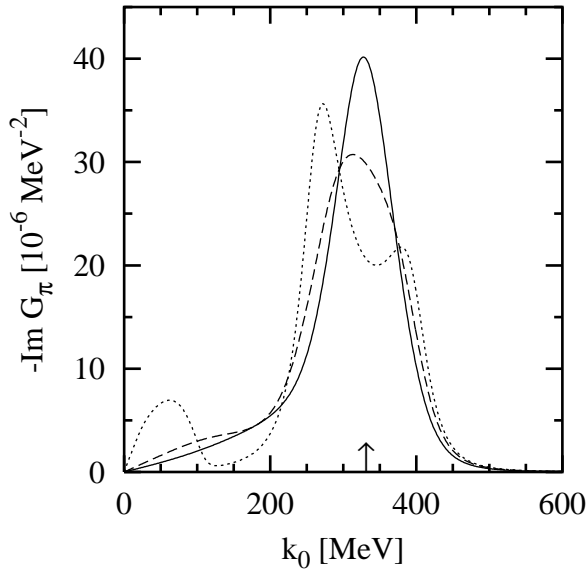,height=7.4cm,width=7.6cm}
\vspace{-.5cm}
\end{center}
\caption{\small Imaginary part of the in-medium pion propagator for fixed
  3-momentum $\abs{\kv} = 300\MeV$ as a function of energy $k_0$, for
  $\rhoN=0.55\rhoo=0.088\fm^{-3}$ and various temperatures: $T=0$
  ($\mu_B=963\MeV$, dotted line), $T=80\MeV$ ($\mu_B=766\MeV$, dashed line)
  and $T=150\MeV$ ($\mu_B=452\MeV$, solid line). For comparison, the arrow
  indicates the pole position of a free pion ($k_0\approx 331\MeV$).}
\label{Fig4}
\end{figure}

The influence of temperature on the medium modifications can be summarized as
follows: For fixed baryon density $\rhoB$, the nucleon density $\rhoN$ becomes
smaller at finite temperature, $T$. This reduces the strength of $Nh$ and $\D h$
excitations. Furthermore, as also Delta states are occupied, $\D h$
excitations become suppressed by Pauli blocking, such that the medium effects
are reduced even if $\rhoN$ is kept constant%
\footnote{\label{Footnote1} In reality these two effects might be less
  important, because the role of the suppressed $Nh=NN^{-1}$ and $\D h=\D
  N^{-1}$ excitations is partly taken over by $B_1^*B_1^{*-1}$ and
  $B_2^*B_1^{*-1}$ excitations~\cite{RappWambach}, where $B_1^*$ and $B_2^*$
  denote any baryon resonances ($\D(1232)$, $N^*(1520)$ etc.). Unfortunately
  most of the $\pi B_1^*B_1^*$ and $\pi B_1^*B_2^*$ coupling constants are not
  well known. For $B_2^*B_1^{*-1}$ excitations with quantum numbers of a
  $\r$-meson (``rhosobars'') it has been found that the inclusion of
  higher resonances reduces the suppression effects by $\approx
  40\%$~\cite{RappWambach}, but does not compensate them.
}
. Another temperature effect is the broadening of the $Nh$ and $\D h$ Lindhard
functions due to thermal motion of the nucleons and Deltas. The peaked
structures in the pion spectral function ($Nh$, $\pi$ and $\D h$ peaks), which
exist at zero temperature, become completely washed out at higher temperatures
as indicated in Fig.~\ref{Fig4}.
%
%

At zero temperature the in-medium pion can be described as a mixture of three
quasiparticles, $(Nh)_L$, $\pi$ and $(\D h)_L$. This so-called ``three-level
model'' is equivalent to neglecting the Fermi motion of the nucleons. As a
consequence of the temperature effects described above, it seems to be not
very reasonable to construct such a model also for $T>0$.
%
\section{The $\r$-meson in hot nuclear matter}
%
\label{Sec4}

\begin{figure}[t]
\begin{center}
\epsfig{file=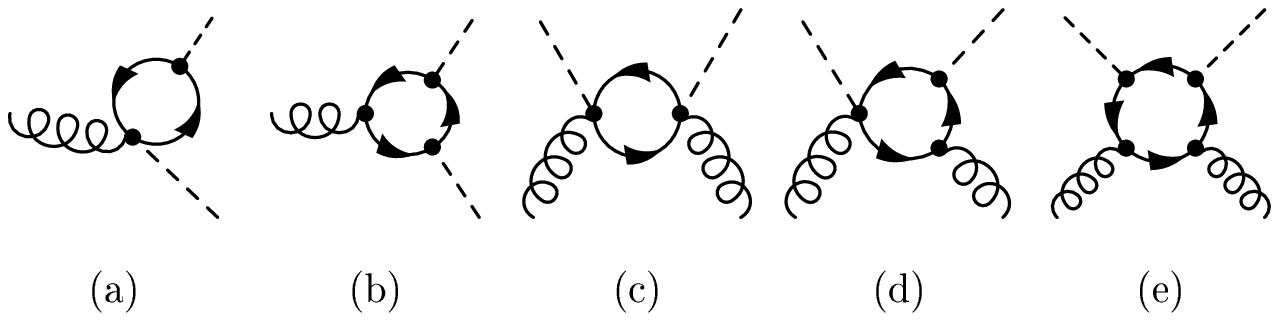}
\vspace{-.5cm}
\end{center}
\caption{\small Some of the $\r\pi\pi$ and $\r\r\pi\pi$ vertex correction
  diagrams corresponding to the $Nh$ pion self-energy.}
\label{Fig5}
\end{figure}

It has been shown by several authors~\cite{KorpaPratt, ChanfraySchuck,
Herrmann}, that simply replacing the free pion propagators in the $\r$-meson
self-energy $\Srpp$ (Figs.~\ref{Fig2}a and \ref{Fig2}b) by in-medium ones
violates gauge invariance and leads to a strong overestimation of the medium
modifications.  To preserve gauge invariance, corrections to the $\r\pi\pi$
and $\r\r\pi\pi$ vertices must be taken into account, which are related to the
pion self-energy through Ward-Takahashi identities. For illustration, some of
the vertex corrections related to the $Nh$ pion self-energy contribution are
shown in Fig.~\ref{Fig5}.
%
%
Neglecting the anomalous magnetic moments of the nucleons, the $\r NN$ and
$\r\pi NN$ couplings can be derived by minimal substitution in the free
nucleon Lagrangian $\calL_N$ and the $\pi N$ interaction Lagrangian
$\calL_{\pi N}$:
\begin{equation}
\calL_{\r N} = -\frac{g}{2}\bar{\psi}\rhoslash\tau_3\psi\ ,\qquad
\calL_{\r\pi N} = ig\frac{f_{\pi NN}}{\mpi} \bar{\psi}\gamma^5\rhoslash
  \vec{\tau}\psi\cdot T_3\vec{\phi}\ .
\end{equation}
In Ref.~\cite{Urban} we have shown that the leading term in a non-relativistic
expansion of the $\r NN$ interaction is the coupling of the $\r$-meson to the
nucleon charge, contributing only to the time component of the $\r NN$ vertex.
The spatial components are of the order $1/\mN$. If we neglect them, only the
diagrams shown in Figs.~\ref{Fig5}a and \ref{Fig5}c are relevant for the
spatial components of the $\r$-meson self-energy%
\footnote{\label{Footnote2} Within a slightly simplified model (zero
  temperature, 2-level model for the pion without Migdal parameter
  $g^\prime$), we have performed the full calculation also allowing for
  $\r$-meson coupling to the convection current. For time-like momenta,
  $q^2>0$, which are relevant for dilepton spectra, the effect of these terms
  is small, but, as expected, in the space-like region near the quasi-elastic
  peak, the approximation becomes quite bad.
}
. For the $\D h$ part, similar vertex corrections must be calculated. This can
be done in analogy to the $Nh$ part described above. The Migdal parameters
$g^\prime$ correspond to an iteration of the $Nh$ and $\D h$ bubbles and
therefore lead to additional vertex corrections. Because of the monopole
form factor $\Gamma_\pi$ at the $\pi NN$ and $\pi N\D$ vertices, the $\r\pi
NN$ and $\r\pi N\D$ vertices must be modified in a non-trivial
way~\cite{Herrmann, Mathiot} as has been described in detail in
Ref.~\cite{Urban}.

\begin{figure}[t]
\begin{center}
\epsfig{file=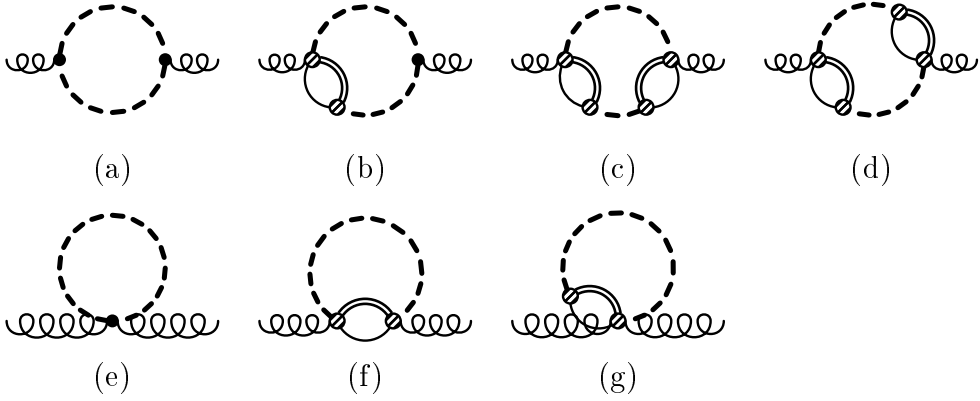}
\vspace{-.5cm}
\end{center}
\caption{\small Feynman diagrams for the spatial components of the pion cloud
  contribution to the $\r$-meson self-energy in hot hadronic matter. The thick
  dashed lines represent in-medium pion propagators, the bubbles represent the
  total pion self-energy with the $\pi NN$ or $\pi N\D$ monopole form factors
  indicated by the hatched blobs.}
\label{Fig6}
\end{figure}

With these vertex corrections the spatial components of the pion cloud
contribution to the $\r$-meson self-energy, $\S_{ij}$ (we omit the index
``$\rho\pi\pi$'' in the following paragraphs), become a sum of seven terms,
corresponding to the seven classes of diagrams shown in Fig.~\ref{Fig6}.
%
%
The diagrams~\ref{Fig6}a and \ref{Fig6}e are the same as in vacuum
(Figs.~\ref{Fig2}a and~\ref{Fig2}b), but with the bare pion propagators
replaced by in-medium ones. The diagrams~\ref{Fig6}b to \ref{Fig6}d and
\ref{Fig6}f to \ref{Fig6}g are obtained by including $\r\pi\pi$ vertex
corrections in diagram~\ref{Fig6}a and $\r\r\pi\pi$ vertex corrections in
diagram~\ref{Fig6}d, respectively.  The diagrams~\ref{Fig6}a to \ref{Fig6}d
and \ref{Fig6}f can be cut into two parts, i.e. they depend on $q_0$ and $\qv$
and contribute to both the real and imaginary parts of $\S_{ij}$. The
diagrams~\ref{Fig6}e and \ref{Fig6}g cannot be cut and are therefore real and
$q_0$-independent.  However, because of the coupling of the $\r$-meson to the
$\pi NN$ and $\pi N\D$ monopole form factor in diagram~\ref{Fig6}g, their
contribution still depends on the 3-momentum $\qv$.

Let us first look at these purely real and energy independent (``constant'')
contributions. Their sum, $\SC_{ij}$, has the structure
\begin{equation}  
\SC_{ij}(\qv)=g^2\int \frac{d^3 k}{(2\pi)^3}
  \sum_{r=5,7}f_{ij}^{(r)}(\qv,\kv)\,J_r(\kv)\ .
\label{SC}
\end{equation}
The $r=5$ and $r=7$ terms correspond to the diagrams~\ref{Fig6}e and
\ref{Fig6}g, respectively. Explicit expressions for the functions
$f_{ij}^{(r)}$, which are simple combinations of $\qv$, $\kv$ and $\Lambda$,
are given in the appendix. The functions $J_5$ and $J_7$ are defined as sums
over Matsubara frequencies $\w_m = \pi m T$:
\begin{equation}
J_5(\kv)=-2T\sum_{m \mbox{\ \scriptsize even}}\calG_\pi(\w_m,\kv)\ ,
\qquad
J_7(\kv)=-2T\sum_{m \mbox{\ \scriptsize even}}
  \calPi(\w_m,\kv)\,\calG_\pi(\w_m,\kv)\ .
\label{J5J7}
\end{equation}
Here $\calG_\pi(\w_m,\kv)$ denotes the imaginary-time pion propagator, which
is related to the retarded propagator $G_\pi$ by its Lehmann
representation~\cite{FetterWalecka},
\begin{equation}
\calG_\pi(\w_m,\kv) = -\frac{1}{\pi}\int d\w\;
  \frac{\Im G_\pi(\w,\kv)}{i\w_m-\w}\ .
\label{Lehmann}
\end{equation}
In an analogous way $\calPi(\w_m,\kv)$ is related to the retarded pion
self-energy $\Pi$. Inserting this into Eq.~(\ref{J5J7}) and using the identity
\begin{equation}
-T\,\lim_{\eta\rightarrow 0}\,\sum_{m \mbox{\ \scriptsize even}}
\frac{e^{i\w_m\eta}}{i\w_m-\w} = \frac{1}{e^{\w/T}-1} =: f(\w)\ ,
\label{Matsubarasum}
\end{equation}
we can evaluate the sum over $\w_m$, with the result
\begin{equation}
J_5(\kv)=-\frac{2}{\pi}\int_0^\infty d\w\;
  \Big(1+2f(\w)\Big)\,\Im G_\pi(\w,\kv)
\label{ReJ5}
\end{equation}
and a similar expression for $J_7$.

Next we turn to the energy-dependent part of $\S_{ij}$, represented by the
diagrams~\ref{Fig6}a to \ref{Fig6}d and~\ref{Fig6}f. It has the structure
\begin{equation}
\S_{ij}(q_0,\qv)-\SC_{ij}(\qv)=g^2\int \frac{d^3 k}{(2\pi)^3}
  \sum_{r=1,2,3,4,6}f_{ij}^{(r)}(\qv,\kv)\,I_r(q_0,\qv,\kv).
\label{SE}
\end{equation} 
The $r=1$, $r=2$ and $r=4$ terms correspond directly to the
diagrams~\ref{Fig6}a,~\ref{Fig6}b and~\ref{Fig6}d, whereas the definitions of
the $r=3$ and $r=6$ terms are somewhat more complicated: The pion self-energy
bubble in diagram~\ref{Fig6}f can be separated into a spin-longitudinal and a
spin-transverse part. The $r=6$ term contains only the transverse part,
whereas the longitudinal part is combined with diagram~\ref{Fig6}c in the
$r=3$ term.

Note that the entire $q_0$ dependence is contained in the functions $I_r$.
These are obtained by analytical continuation of functions $J_r$, which
originally are defined only for discrete Matsubara frequencies $\w_n$, $n$
even:
\begin{equation}
I_r(q_0,\qv,\kv)=J_r(\w_n\rightarrow -i(\qe),\qv,\kv)\ .
\label{Anacont}
\end{equation}
For example, the function $J_1$ is defined as follows:
\begin{equation}
J_1(\w_n,\qv,\kv)=-2T\sum_{m \mbox{\ \scriptsize even}}
  \calG_\pi(\w_m,\kv)\,\calG_\pi(\w_{m+n},\kv+\qv)\ .
\label{J1}
\end{equation}
The definitions of the other functions $J_r$ are listed in the appendix.

As described already for $J_5$, we insert the Lehmann representation of
$\calG_\pi$ into Eq.~(\ref{J1}) and evaluate the sum over $\w_m$ by using
Eq.~(\ref{Matsubarasum}). After these steps the analytical continuation of
$J_1$ to $I_1$ reduces to a simple replacement according to
Eq.~(\ref{Anacont}). For the imaginary part we obtain
\begin{eqnarray}
\Im I_1(q_0,\qv,\kv)&=&
  -\frac{2}{\pi}\int_0^{q_0} d\w\;\Big(1+f(\w)+f(q_0-\w)\Big)
\nonumber \\*
&&\hspace{2cm}\times\,
  \Im G_\pi(\w,\kv)\,\Im G_\pi(q_0-\w,\kv+\qv)
\nonumber \\
&&-\frac{2}{\pi}\int_0^\infty d\w\;\Big(f(\w)-f(q_0+\w)\Big)
\nonumber \\*
&&\hspace{2cm}\times
  \Big(\Im G_\pi(\w,\kv)\,\Im G_\pi(q_0+\w,\kv+\qv)
\nonumber \\*
&&\hspace{2.25cm}
  +\Im G_\pi(q_0+\w,\kv)\,\Im G_\pi(\w,\kv+\qv)\Big)\ .
\label{ImI1}
\end{eqnarray}
In the same way we derive similar expressions for the imaginary parts of the
remaining functions $I_2$ to $I_4$ and $I_6$. Rather than by direct
calculation, the real part of the energy-dependent piece of the $\r$-meson
self-energy, $\S_{ij}-\SC_{ij}$, can be computed more efficiently from the
imaginary part using a dispersion relation. The convergence of the dispersion
integral is improved by subtracting the vacuum contribution:
\begin{equation}
\Re\S_{ij}(q_0,\qv)=\Re\Svac_{ij}(q_0,\qv)+\SCmed_{ij}(\qv)
  -\frac{1}{\pi}\,{\mathcal P}\!\!\int_0^{\infty}d\w^2\,
  \frac{\Im\Smed_{ij}(\w,\qv)}{q_0^2-\w^2}\ .
\end{equation}

\begin{figure}[p]
\begin{center}
\epsfig{file=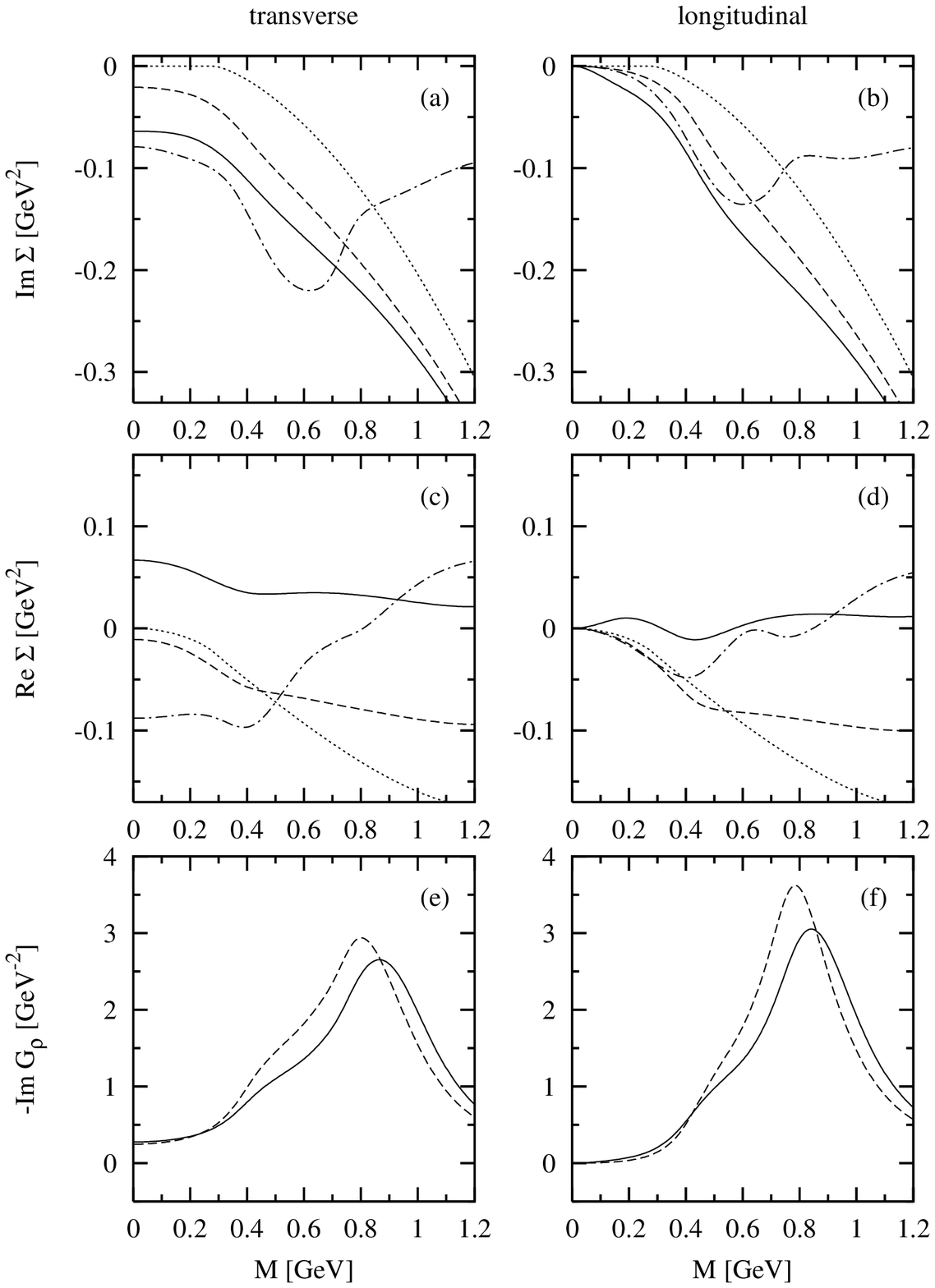}
\vspace{-.5cm}
\end{center}
\caption{\small (a) to (d) Transverse and longitudinal $\r$-meson
  self-energies for fixed 3-momentum $\abs{\qv}=300\MeV$ as functions of the
  invariant mass $M$ in vacuum (dotted lines) and for
  $\rhoN=0.55\rhoo=0.088\fm^{-3}$: $\Srpp$ for $T=0$
  ($\mu_B=963\MeV$, dashed lines), $\Srpp$ for $T=150\MeV$ ($\mu_B=452\MeV$,
  solid lines) and $\SrB+\SrM$ for $T=150\MeV$ (dashed-dotted lines). (e) and
  (f) $\r$-meson spectral functions for $\rhoN=0.55\rhoo=0.088\fm^{-3}$ and
  $T=150\MeV$ within the full model (solid lines) and neglecting temperature
  effects in $\Srpp$ (dashed lines).}
\label{Fig7}
\end{figure}

Figs.~\ref{Fig7}a to~\ref{Fig7}d display numerical results for the $\r$-meson
self-energy in a baryonic medium with $\rhoN=0.55\rhoo$.  Because of the
approximations employed for the $\r NN$ vertex, we restrict ourselves to
$q^2>0$ (see footnote on page~\pageref{Footnote2}) and plot the results as
functions of the invariant mass $M=\wurzel{q^2}$.
%
%

Let us first discuss the imaginary part of the self-energy (Figs.~\ref{Fig7}a
and~\ref{Fig7}b). The dashed lines show the transverse and longitudinal parts
of $\Srpp$ for $\rhoN=0.55\rhoo$ and $T=0$. Compared with $\Svac$ (dotted
lines), the following medium modifications are visible: The threshold
disappears, because the $\r$-meson can ``decay'' into two $Nh$ excitations.
The steep rise of $\Im\Srpp$ at $M\approx 400\MeV$ and the large imaginary
part above this energy is mainly a result of the ``decay'' of the $\r$-meson
into a pion and a transverse $\D h$ excitation ($r=6$ term,
diagram~\ref{Fig6}f). A more detailed discussion of these effects can be found
in Ref.~\cite{Urban}.

The solid lines show $\Srpp$ for $T=150\MeV$. For a discussion of the
differences to the $T=0$ case, consider e.g. Eq.~(\ref{ImI1}), which describes
the imaginary part of the $r=1$ term. At higher energies ($M\gtrsim 300\MeV$)
the imaginary part is dominated by the first integral, which contributes
already at $T=0$, but is enhanced at $T>0$ by the Bose-factor
$1+f(\w)+f(q_0-\w)$. It describes the ``decay'' of the $\r$-meson into two
in-medium pions with energies $\w$ and $q_0-\w$ (Fig.~\ref{Fig2}c with the
free pions replaced by in-medium ones: $\r\rightarrow\pimed+\pimed$, including
processes such as $\r\rightarrow Nh+\pi$, $\r\rightarrow\D h+\pi$ etc.). At
very low energies ($M\lesssim 200\MeV$), the second integral in
Eq.~(\ref{ImI1}) is more important. It exists only at $T>0$ because of the
factor $f(\w)-f(q_0+\w)$. Therefore $\Im\Srpp$ increases strongly with
temperature in this low-energy region. The corresponding physical process is
the ``absorption'' of the $\r$-meson by a thermal pion with energy $\w$,
giving the outgoing pion the energy $q_0+\w$ (Fig.~\ref{Fig2}d with the free
pions replaced by in-medium ones: $\r+\pimed\rightarrow\pimed$). In the
absence of baryons, $\rhoB=0$, this is possible only for space-like 4-momenta,
$q^2<0$, i.e. for $q_0<\abs{\qv}$ (see Section~\ref{Sec2}). At finite baryon
density the situation changes, however, as a consequence of the broadening of
the pion spectral function, such that this term contributes also for $q^2>0$
(e.g. via the process $\r+\pi\rightarrow\D h$, which is of course contained in
the class of processes $\r+\pimed\rightarrow\pimed$.)

Finally, the dashed-dotted curves represent the sum of the remaining
self-energy contributions, $\SrB+\SrM$, for $T=150\MeV$ and $\rhoN=0.55\rhoo$.
The contribution $\SrM$ has been considered already in Section~\ref{Sec2}, we
have taken it from Ref.~\cite{RappGale} and displayed it in Fig.~\ref{Fig3}.
The contribution $\SrB$ corresponds to diagrams as shown in Fig.~\ref{Fig1}b.
It is described in Ref.~\cite{RappUrban} for $T=0$ and has been extended to
finite temperature~\cite{RappChanfrayWambach, RappWambach}. The parameters of
the model have been fixed by fitting $B\rightarrow\r N$ decay widths and
photoabsorption cross sections of the nucleon and of nuclei~\cite{RappUrban}.
(Note that in Ref.~\cite{RappWambach} the photoabsorption cross-section is
fitted with a very soft $\pi NN$ form factor, $\L=300\MeV$, while we take
$\L=550\MeV$ as in Ref.~\cite{RappUrban}. Therefore in Ref.~\cite{RappWambach}
$\SrB$ is larger and $\Srpp$ is smaller than in this article.)

Let us now turn to the real part of the self-energy (Figs.~\ref{Fig7}c
and~\ref{Fig7}d). Comparing the dotted, dashed and solid curves, we find that
both, density and temperature, increase $\Re\Srpp$ in the region of the
$\r$-meson pole, shifting it to higher energies. For the case $\rhoB=0$ we
have noted this temperature effect already in Section~\ref{Sec2}
(Fig.~\ref{Fig3}). It is caused mainly by $\SC$ (factor $1+2f(\w)$ in
Eq.~(\ref{ReJ5})) and by the energy-dependent real parts related to the
second integral in Eq.~(\ref{ImI1}). Contrary to $\Re\Srpp$, the real part of
$\SrB+\SrM$ (dashed-dotted lines) is almost zero at $M\approx 800\MeV$ and
therefore does not shift the position of the pole.

The medium effects can be seen more clearly in the spectral functions than in
the self-energies. They are displayed in Figs.~\ref{Fig7}e and~\ref{Fig7}f.
The solid lines represent the full calculation for $T=150\MeV$ and
$\rhoN=0.55\rhoo$, i.e. using Eq.~(\ref{Grho}) with $\Stot=\Srpp+\SrB+\SrM$.
As we expected already from the discussion of the self-energies, the width of
the $\r$-meson is strongly increased compared with the vacuum, and the pole
mass, i.e. the position of the maximum, is shifted upwards from $770\MeV$ in
vacuum to $865\MeV$ for $\Im G_{\r\,T}$ and to $840\MeV$ for $\Im G_{\r\,L}$,
respectively. Since we are mainly interested in the temperature effects in
$\Srpp$, we show as the dashed curve a spectral function that has been
obtained by neglecting these temperature effects, i.e. $\Stot$ has been
replaced by $\Srpp(T=0)+\SrB+\SrM$. Now one can see clearly that about $70\%$
of the total mass shift is due to the temperature effects in $\Srpp$.
%
\section{Dilepton and photon rates}
%
\label{Sec5}
Our main motivation for the investigation of the $\r$-meson in hot and dense
matter is the description of dilepton production in heavy-ion collisions.
Assuming hadronic matter in (local) thermal and chemical equilibrium
(temperature $T$, baryon chemical potential $\mu_B$) and Sakurai's vector
dominance model (VDM)~\cite{Sakurai}, the production rate ($R=dN/d^4x$) of
$e^+e^-$ pairs with 4-momentum $q$ in the low invariant mass region ($M =
\wurzel{q^2} < 1 \GeV$) in the isovector ($\r$-meson) channel is given
by~\cite{ChanfraySchuck}
\begin{equation}
\frac{dR_{e^+e^-}}{d^4q}=\frac{\alpha^2}{3\pi^3q^2}\,
  \frac{1}{e^{q_0/T}-1}\,\frac{m_{\r}^{(0)\,4}}{g^2}\,
  g_{\mu\nu}\,\Im G_\r^{\mu\nu}(q_0,\qv;\mu_B,T)\ ,
\label{DPR}
\end{equation}
where $\alpha=e^2/(4\pi)=1/137$. In the derivation of this expression the
electron mass has been neglected. However, if one includes the baryonic
contribution $\SrB$, it is necessary to formulate the VDM such that the $\r
NB$ and $\gamma NB$ couplings can be adjusted
independently~\cite{KrollLeeZumino, FrimanPirner}. Then the expression
$m_{\r}^{(0)\,4}/g^2\,g_{\mu\nu}\,\Im G_\r^{\mu\nu}$ in Eq.~(\ref{DPR}) is
replaced by~\cite{RappWambach}
\begin{equation}
\frac{m_{\r}^{(0)\,4}}{g^2}\,g_{\mu\nu}\,\Im G_\r^{\mu\nu}\rightarrow
\frac{1}{g^2}(2F_T+F_L)\ ,
\label{FTFL}
\end{equation}
where
\begin{eqnarray}
F_L&=&-m_{\r}^{(0)\,4}\,\Im G_{\r\,L}\ ,
\label{FL}\\
F_T&=&-(\Im\SrppT+\Im\SrMT)\,\abs{d_{\r}-1}-\Im\SrBT\,\abs{d_{\r}-r_B}\ ,
\label{FT}\\
d_{\r}&=&(q^2-\SrppT-\SrMT-r_B \SrBT)\,G_{\r\,T}\ .
\end{eqnarray}
The parameter $r_B=0.7$~\cite{RappUrban} is the ratio of the actual $\gamma
NB$ coupling to the $\gamma NB$ coupling derived in Sakurai's VDM.

\begin{figure}[t]
\begin{center}
\epsfig{file=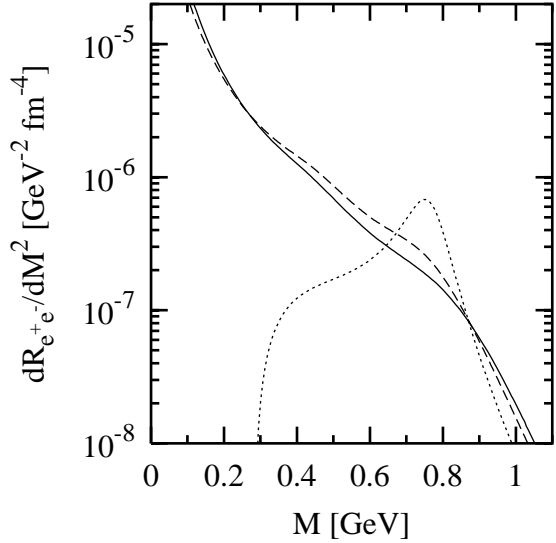,height=7.06cm,width=7.62cm}
\vspace{-.5cm}
\end{center}
\caption{\small Dilepton production rate $dR_{e^+e^-}/dM^2$ for
  $T=150\MeV$ obtained without medium modifications of the $\r$-meson (dotted
  line) and with medium modifications for $\rhoN=0.55\rhoo$: full model
  (solid line) and neglecting temperature effects in $\Srpp$ (dashed line).}
\label{Fig8}
\end{figure}

Results for the dilepton rate as a function of $M$, i.e. integrated over
$\qv$, are displayed in Fig.~\ref{Fig8}
%
%
for $T=150\MeV$ and $\rhoN=0.55\rhoo$, corresponding to conditions as expected
to be realized in the $E_{\mathit{lab}}=40\,\mathrm{AGeV}$ run at the
CERN-SpS. Without any medium modifications of the $\r$-meson, the dilepton
rate is peaked around $M=770\MeV$ (dotted line). With medium modifications,
the peak disappears and the dilepton rate for $M\lesssim 650\MeV$ is strongly
enhanced (solid line).  If one neglects temperature effects in $\Srpp$ (dashed
line), the rate in this region is slightly overestimated (by $15\%$ at
$M=400\MeV$) because of the $\r$-meson mass shift observed in
Section~\ref{Sec4}. The difference is even larger at $M\approx 770\MeV$,
where the full result (solid line) is $30\%$ below the result obtained by
neglecting temperature effects in $\Srpp$ (dashed line). Therefore the
temperature effects in $\Srpp$, principally the mass shift, might improve the
description of the experimental dilepton spectra which, in this invariant mass
region, are almost saturated by $e^+e^-$-pairs expected from free $\w$-meson
decays after freeze-out.

Besides dilepton pairs, also direct real photons (i.e. real photons produced
inside the fireball) can be used as probes for the hot and dense phase of the
fireball in heavy-ion experiments. If we make the same assumptions as in the
derivation of Eq.~(\ref{DPR}), the rate for direct real photon emission is
given by~\cite{SteeleYamagishiZahed}
\begin{equation}
q_0\frac{dR_\gamma}{d^3q}=\frac{\alpha}{2\pi^2}\,
  \frac{1}{e^{q_0/T}-1}\,\frac{m_{\r}^{(0)\,4}}{g^2}\,
  g_{\mu\nu}\,\Im G_\r^{\mu\nu}(q_0,\qv;\mu_B,T)\ ,
\end{equation}
where $q_0=\abs{\qv}$ is the energy of a photon with momentum $\qv$. Again,
when $\SrB$ is included, the term $m_{\r}^{(0)\,4}/g^2\,g_{\mu\nu}\,\Im
G_\r^{\mu\nu}$ must be replaced according to Eq.~(\ref{FTFL}), but since $F_L$
vanishes for $q^2=0$, only the transverse part $F_T$ contributes.

\begin{figure}[t]
\begin{center}
\epsfig{file=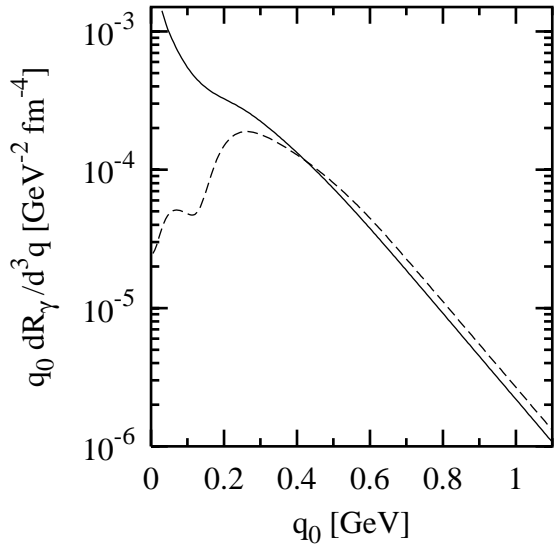,height=7.06cm,width=7.62cm}
\vspace{-.5cm}
\end{center}
\caption{\small Direct photon production rate $q_0\,dR_{\gamma}/d^3q$ for
  $\rhoN=0.55\rhoo$ and $T=150\MeV$ within the full model (solid line) and
  neglecting temperature effects in $\Srpp$ (dashed line).}
\label{Fig9}
\end{figure}

Numerical results are shown in Fig.~\ref{Fig9}.
%
%
Again the solid line represents the full calculation, while for the dashed
line the temperature dependence of $\Srpp$ for fixed $\rhoN$ has been
neglected. The enhancement of the solid curve below $300\MeV$ is a consequence
of the imaginary part of the ``absorption'' terms discussed at the end of
Section~\ref{Sec4}, which in the context of photon emission correspond to
processes such as $\pimed\rightarrow\pimed+\r\rightarrow\pimed+\gamma$, e.g.
$\D h\rightarrow\pi+\r\rightarrow\pi+\gamma$.
%
\section{Summary and conclusions} 
%
\label{Summary}
In a previous article~\cite{Urban} we have extended the models from
Refs.~\cite{ChanfraySchuck, Herrmann}, describing the modifications of the
$\r$-meson through the in-medium pion propagator in cold nuclear matter, i.e.
through the self-energy $\Srpp$, to finite 3-momentum of the $\r$-meson.
In the present work we have also included finite-temperature effects by
evaluating both, the pion and $\r$-meson self-energy diagrams within the
imaginary-time formalism. We stress that this is the first consistent
finite-temperature calculation of $\Srpp$, whereas the medium modifications
through resonant $\r$-meson scattering off baryons ($\SrB$) or mesons ($\SrM$)
at $T>0$ have already been evaluated in Refs.~\cite{RappChanfrayWambach,
RappWambach, RappGale}.

Already at moderate nucleon densities, $\rhoN=0.55\rhoo$, the self-energy
contribution $\Srpp$ leads to significant changes of the $\r$-meson spectral
function. Let us summarize the most important effects: First, the width of the
$\r$-meson is increased. Second, the maximum of the $\r$-meson spectral
function, i.e. the $\r$-meson ``mass'', is shifted upwards in the medium.  Of
course, the other contributions, $\SrB$ and $\SrM$, result in an additional
broadening of the spectral function, but they produce almost no mass shift.
Concentrating on the medium modifications through $\Srpp$, the $\r$-meson
width and mass increase with both, density and temperature. However, the
broadening depends predominantly on the density, whereas the mass shift is
mainly a temperature effect.

The broadening of the spectral function results in a strong enhancement of the
dilepton production rate for invariant masses below $\approx 650\MeV$, while
the peak at $770\MeV$, which one would expect in a naive model (without medium
modifications of the $\r$-meson), disappears. The mass shift from the
temperature effects in $\Srpp$ helps to suppress the dilepton rate around
$770\MeV$ further, but also reduces the enhancement below $650\MeV$ from the
broadening a little. However, the results shown in Sect.~\ref{Sec5} might be
quantitatively somewhat modified if one applies the soft $\pi NN$ form factor
of $\L=300\MeV$, as in Ref.~\cite{RappWambach}, instead of $\L=550\MeV$.

In some respects the model could be further improved. First, for the
description of the pion in nuclear matter also higher resonances beyond the
$\D(1232)$ should be included (see footnote on page~\pageref{Footnote1}).
Second, for the case $T>0$ but $\rhoB=0$, it has been shown that in a
consistent treatment of $\r$- and $a_1$-mesons, respecting chiral symmetry,
the leading temperature effect essentially consists of a mixing of vector and
axial-vector correlators, i.e. $\r$- and $a_1$-meson
propagators~\cite{DeyEletsky}. To relate the measured dilepton spectra to
potential signals of chiral symmetry restoration, it is furthermore mandatory
to study the medium modifications of $\r$- and $a_1$-mesons within a chiral
symmetric model also at $\rhoB\neq 0$. Work in this direction is in progress.
\vspace{1cm}
\begin{center}
%
\bf\large Acknowledgements 
%
\end{center}
This work was supported in part by GSI, BMBF and NSF grant NSFPHY98-00978.
One of us (RR) is supported by DOE grant no. DE-FG02-88ER40388 and is
grateful for the hospitality at TU Darmstadt.
\vspace{1cm}
%
%
\begin{appendix}
\section{Explicit expressions and definitions}
In Section~\ref{Sec4} only the structure of the formula for the spatial
components of the $\r$-meson self-energy $\Srpp\equiv\S$ has been given
(Eqs.~(\ref{SC}) and~(\ref{SE})). The explicit expression reads
\begin{eqnarray}
\S_{ij}(q_0,\qv)&=&
  \frac{g^2}{4}\int \frac{d^3k}{(2\pi)^3}\bigg(
  (2k+q)_i(2k+q)_j
  \,I_1(q_0,\qv,\kv)
  \nonumber\\
&&+4\frac{(k_i(\L^2-\kv^2)-q_i\kv^2)(2k+q)_j}{\L^2+(\kv+\qv)^2}
  \,I_2(q_0,\qv,\kv)
  \nonumber\\
&&+2\frac{(k_i(\L^2-\kv^2)-q_i\kv^2)(k_j(\L^2-\kv^2)-q_j\kv^2)}
  {\kv^2(\L^2+(\kv+\qv)^2)^2}
  \,I_3(q_0,\qv,\kv)
  \nonumber\\
&&+2\frac{(k_i(\L^2-\kv^2)-q_i\kv^2)(k_j(\L^2-(\kv+\qv)^2)+q_j\L^2)}
  {(\L^2+(\kv+\qv)^2)(\L^2+\kv^2)}
  \,I_4(q_0,\qv,\kv)
  \nonumber\\
&&+2\delta_{ij}
  \,J_5(\kv)
  \nonumber\\
&&+2\frac{(\L^2+\kv^2)^2}{(\L^2+(\kv+\qv)^2)^2}
  \Big(\delta_{ij}-\frac{k_ik_j}{\kv^2}\Big)
  \,I_6(q_0,\qv,\kv)
  \nonumber\\
&&-4\Big(\frac{\kv^2\delta_{ij}}{\L^2+\kv^2}
  +\frac{(2k+q)_i(k_j(\L^2-\kv^2)-q_j\kv^2)}
  {(\L^2+(\kv+\qv)^2)(\L^2+\kv^2)}\Big)
  \,J_7(\kv)
  \bigg)
  \nonumber\\
&&+\,(i\longleftrightarrow j)\ .
\end{eqnarray}

The retarded functions $I_r$ are given by analytical continuation of the
corresponding imaginary-time functions $J_r$ to the real axis according to
Eq.~(\ref{Anacont}). The functions $J_1$ to $J_7$ are defined as follows:
\begin{eqnarray}
J_1(\w_n,\qv,\kv)&=&-2T\sum_{m \mbox{\ \scriptsize even}}
  \calG_\pi(\w_m,\kv)\,\calG_\pi(\w_{m+n},\kv+\qv)\ ,
\\
J_2(\w_n,\qv,\kv)&=&-2T\sum_{m \mbox{\ \scriptsize even}}
  \calPi(\w_m,\kv)\,\calG_\pi(\w_m,\kv)\,
  \calG_\pi(\w_{m+n},\kv+\qv)\ ,
\\
J_3(\w_n,\qv,\kv)&=&-2T\sum_{m \mbox{\ \scriptsize even}}
  \calPi_L(\w_m,\kv)\,\calG_\pi(\w_{m+n},\kv+\qv)\ ,
\\
J_4(\w_n,\qv,\kv)&=&-2T\sum_{m \mbox{\ \scriptsize even}}
  \calPi(\w_m,\kv)\,\calG_\pi(\w_m,\kv)
\nonumber \\*
&&\hspace{2cm}\times
  \calPi(\w_{m+n},\kv+\qv)\,\calG_\pi(\w_{m+n},\kv+\qv)\ ,
\\
J_5(\kv)&=&-2T\sum_{m \mbox{\ \scriptsize even}}\calG_\pi(\w_m,\kv)\ ,
\\
J_6(\w_n,\qv,\kv)&=&-2T\sum_{m \mbox{\ \scriptsize even}}
  \calPi(\w_m,\kv)\,\calG_\pi(\w_{m+n},\kv+\qv)\ ,
\\
J_7(\kv)&=&-2T\sum_{m \mbox{\ \scriptsize even}}
  \calPi(\w_m,\kv)\,\calG_\pi(\w_m,\kv)\ .
\end{eqnarray}

The imaginary-time pion self-energy $\calPi$ and propagator $\calG_\pi$ are
related to the corresponding retarded functions $\Pi$ and $G_\pi$ by their
Lehmann representations
\begin{equation}
\calPi(\w_m,\kv) = -\frac{1}{\pi}\int d\w\;
  \frac{\Im \Pi(\w,\kv)}{i\w_m-\w}\ ,\quad
\calG_\pi(\w_m,\kv) = -\frac{1}{\pi}\int d\w\;
  \frac{\Im G_\pi(\w,\kv)}{i\w_m-\w}\ .
\end{equation}
As a consequence, such representations exist also for $\calPi\,\calG_\pi$ and
$\calPi_L$, which are combinations of these functions. The longitudinal
spin-isospin response function $\Pi_L$ is defined by
\begin{eqnarray}
\Pi_L(k_0,\kv)
&=&\Pi(k_0,\kv)+\Pi(k_0,\kv)\,\kv^2\,G_\pi(k_0,\kv)\,\Pi(k_0,\kv)
\nonumber\\*
&=&(k^2-\mpi^2)\,\Pi(k_0,\kv)\,G_\pi(k_0,\kv)\ .
\end{eqnarray}
\end{appendix}
%
%

%

\newpage
\begin{thebibliography}{99}
\bibitem{Shuryak} E.V. Shuryak, Phys. Lett. B 78 (1979) 150.
\bibitem{Sakurai} J.J. Sakurai, Ann. Phys. 11 (1960) 1.
\bibitem{KrollLeeZumino} N.M. Kroll, T.D. Lee and B. Zumino,
  Phys. Rev. 157 (1967) 1376.
\bibitem{RappWambach1} R. Rapp and J. Wambach, preprint hep-ph/9909229,
  to be published in Adv. Nucl. Phys.
\bibitem{EricsonWeise} T. Ericson and W. Weise, Pions and Nuclei
  (Oxford University Press, New York, 1988).
\bibitem{GaleKapusta87} C. Gale and J. Kapusta, Phys. Rev. C 35 (1987) 2107.
\bibitem{KorpaPratt} C.L. Korpa and S. Pratt, Phys. Rev. Lett. 64 (1990) 1502.
\bibitem{ChanfraySchuck} G. Chanfray and P. Schuck,
  Nucl. Phys. A 555 (1993) 329.
\bibitem{Herrmann} M. Herrmann, B. Friman and W. N\"orenberg, 
  Nucl. Phys. A 560 (1993) 411;\\
  M. Herrmann, PhD thesis, Darmstadt 1992 (GSI-Report 92-10).
\bibitem{FrimanPirner} B. Friman and H.J. Pirner,
  Nucl. Phys. A 617 (1997) 496.
\bibitem{PetersPost} W. Peters, M. Post, H. Lenske, S. Leupold and U. Mosel,
  Nucl. Phys. A 632 (1998) 109.
\bibitem{CERES} G. Agakichiev et al., CERES collaboration, 
  Phys. Rev. Lett. 75 (1995) 1272;\\
  P. Wurm for the CERES collaboration, Nucl. Phys. A 590 (1995) 103c.\\
  G. Agakichiev et al., CERES collaboration, Phys. Lett. B 422 (1998) 405. 
\bibitem{RappChanfrayWambach} R. Rapp, G. Chanfray and J. Wambach,
  Nucl. Phys. A 617 (1997) 472.
\bibitem{RappWambach} R. Rapp and J. Wambach, preprint hep-ph/9907502.
\bibitem{Urban} M. Urban, M. Buballa, R. Rapp and J. Wambach,
  Nucl. Phys. A 641 (1998) 433.
\bibitem{RappUrban} R. Rapp, M. Urban, M. Buballa and J. Wambach,
  Phys. Lett. B 417 (1998) 1.
\bibitem{FetterWalecka} A.L. Fetter and J.D. Walecka, Quantum Theory
  of many-particle systems (McGraw-Hill, New York, 1971). 
\bibitem{GaleKapusta91} C. Gale and J.I. Kapusta,
  Nucl.Phys. B 357 (1991) 65.
\bibitem{RappGale} R. Rapp and C. Gale, Phys. Rev. C 60 (1999) 24903. 
\bibitem{Migdal} A.B. Migdal, Rev. Mod. Phys. 50 (1978) 107.
\bibitem{Mathiot} J.F. Mathiot, Nucl. Phys. A 412 (1984) 201.
\bibitem{SteeleYamagishiZahed} J.V. Steele, H. Yamagishi and I. Zahed,
  Phys. Rev. D 56 (1997) 5605. 
\bibitem{DeyEletsky} M. Dey, V.L. Eletsky and B.L. Ioffe,
  Phys. Lett. B 252 (1990) 620,\\
  V.L. Eletsky and B.L. Ioffe, Phys. Rev. D 51 (1995) 2371.
\end{thebibliography}
\end{document}